# Cooperative ISP Traffic Shaping Schemes in Broadband Shared Access Networks

Luke Farmer and Kyeong Soo Kim, *Member, IEEE*
College of Engineering, Swansea University
Swansea, SA2 8PP, United Kingdom
Email:{525709, k.s.kim}@swansea.ac.uk

*Abstract*—Traffic shaping is a mechanism used by Internet Service Providers (ISPs) to limit subscribers' traffic based on their service contracts. This paper investigates the current implementation of traffic shaping based on the token bucket filter (TBF), discusses its advantages and disadvantages, and proposes a cooperative TBF that can improve subscribers' quality of service (QoS)/quality of experience (QoE) without compromising business aspects of the service contract model by proportionally allocating excess bandwidth from inactive subscribers to active ones based on the long-term bandwidths per their service contracts.

## I. INTRODUCTION

RECENT studies on the effects of shaping the traffic from subscribers[1] by Internet service providers (ISPs) in access networks have shown the advantages of exploiting *bursty* nature of traffic using a token bucket filter (TBF) with a large-size token bucket in a short time scale (e.g., seconds) [1][2].

Traffic shaping in access networks is commonly construed as a negative; limiting traffic based on service contracts, time of day, or type of traffic (e.g., throttling peer-to-peer applications) via protocol or port inspection. But there exist positive traffic shaping methods that improve the quality of service (QoS) for some traffic types, e.g., allowing delay-sensitive applications to pass their packets through a system without queuing [3].

This paper focuses on the TBF scheme used in current ISP traffic shaping and studies on ways to improve its performance in shared access networks in a long time scale (e.g., minutes to hours), too. Our proposals use the TBF as a base for a rate sharing system that increases average rates for active subscribers without compromising ISP business or QoS/quality of experience (QoE) of inactive subscribers.

The rest of the paper is organized as follows: Section II provides a detailed review of the TBF scheme currently used in ISP traffic shaping. Section III describes a new traffic shaping scheme as an extension of TBF. Section IV presents preliminary results of simulation experiments in a realistic environment to show the benefits of the proposed scheme and to find the circumstances in which it is most effective. Section V concludes our work in this paper.

## II. TOKEN BUCKET FILTER (TBF)

Describing the proposed system requires a brief explanation of the inner workings of the original TBF. The TBF takes advantage of the short-term burstiness of typical Internet traffic — e.g., the few seconds between hypertext transfer protocol (HTTP) requests while browsing the web, variable bit rate (VBR) video streaming — and allows subscribers to temporarily transmit at much higher rates than they are assigned (called a burst) by saving up tokens while their connection is being used at less than the assigned rate (i.e., token generation rate). Over a long period of time, a subscriber's average rate would be equal to or less than its assigned rate, yet the short-term perceived performances could be improved. This system can be used either to help guarantee a minimum QoS [4] or as an advantageous, marketable feature [2].

### A. Operations

The TBF as traffic shaper is positioned between the egress classifier and the scheduler in an access switch and uses a set of first-in-first-out (FIFO) queues, one for each subscriber. Packets entering the TBF carry an identification number assigned by the egress classifier, which is used to select the appropriate queue [1]. Each queue works independently and has its own set of shaping parameters: Token generation rate (equal to the downstream transfer rate assigned to this subscriber by the ISP), peak rate (the maximum rate which any single subscriber can transmit at; normally depends on the underlying physical-layer technology, but may be configured [2]), and a data structure that contains the current burst length available to this subscriber (i.e., token bucket) which has a maximum size.

Tokens are constantly added to the bucket at a token generation rate. When a packet enters the queue, its length is compared to the current amount of tokens in the bucket. If it conforms (i.e., packet length is less than or equal to the current amount of tokens in the bucket), then the packet is sent to the scheduler and the amount of tokens equal to the packet length is removed from the token bucket. If the current amount of tokens allows for it, therefore, multiple packets can be processed at a peak rate — many times higher than the token generation rate — until there are not enough tokens, and the transfer rate returns to the assigned token generation rate. If a packet does not conform, it must wait in the queue until enough tokens are available [5].

---

[1] A subscriber could be made up of multiple users/hosts, but from the ISP's perspective, a subscriber is seen as a single entity.

Shaping the peak rate is performed the same way. A second, much smaller token bucket equal to the size of the maximum transmission unit (MTU) is also checked for packet conformity. A packet must conform to both TBFs before it can be moved on to the scheduler.

The effect of this system on user-perceived performances has been studied in detail in [1], where it was concluded that, provided a large token bucket size and reasonably small number of subscribers sharing a single access link, user-perceived performances in terms of both response and throughput is greatly improved, particularly with intensive transmissions such as file transfer protocol (FTP) traffic.

*B. Issues in the Current Scheme*

While the benefits of TBF are substantial, the finite size of a token bucket and the independent operation of each queue create room for further improvement. Note that, when a token bucket is full, any further tokens generated for that subscriber are discarded irrespective of the current network status. This could result in a rather large amount of waste in the usage of network resources; even when there are only a few active subscribers in a shared access network and thereby there is plenty of excess bandwidth from inactive subscribers, the maximum sustained throughputs of active subscribers are still limited by their assigned token generation rates.

The situation is not going to improve at all because, with the development of new access technology offering higher line rates and the increasing use of peer-to-peer file sharing, it will be more common for users to attempt to download very large files relative to their assigned rates [6]. The TBF helps little in this scenario because the current implementation cannot take any advantage of excess bandwidth from inactive subscribers.

III. COOPERATIVE TOKEN BUCKET FILTER (CTBF)

The main principle of the proposed system is to remove the conceptual barrier between queues within a single shared access switch and enable them to share information via a controller, allowing the use of token generation rate sharing algorithms that significantly improve actual performances during the downtime of neighboring subscribers. This increased performance gives active users longer bursts, allowing them to complete large transmissions more quickly, reducing the chance of congestion occurring that has a negative impact on QoS, all without any perceptible difference in QoS for those less active subscribers.

*A. Backgrounds*

Before describing the proposed system, it must be made clear that the business of ISP service contracts is a factor in the design of any positive traffic shaping scheme as touched on in [2]. ISPs want their customers to upgrade to more expensive services, with the incentive being higher and/or more stable rates. With TBF, because queues are independent, this is not a problem as the average rate of each subscriber can easily be controlled. However, if every discarded token is given to other subscribers regardless of their services, this may result in a subscriber with a cheap service receiving extremely large rate bonus (e.g., several or tens of times faster than its assigned rate) depending on the status of other subscribers, almost completely removing any incentive for them to upgrade.

The solution is a proportional distribution and (optionally) a proportional cap on the rate bonus that one subscriber can receive. This way, if a subscriber decides to stay with their cheap service, the bonus rate they receive (from individual subscribers, not the maximum bonus) will gradually decline as other subscribers upgrade their own services. Provided that the subscriber is made aware of how this distribution works and the exact benefit of upgrading, it gives them a strong incentive to upgrade sooner rather than later. And because the bonus rate is always greater than what the subscriber is paying for, this cannot be considered an unethical business practice.

*B. Operations*

When the amount of tokens in a token bucket becomes less than a configurable threshold (e.g., 95% of the token bucket size), that subscriber is declared inactive. The threshold value is used mainly to prevent smaller packet sequences (e.g., HTTP traffic) from flipping a subscribers' active/inactive state unnecessarily. But for the purpose of keeping the examples in this document simple, this threshold is ignored in the following.

The cooperative token bucket filter (CTBF) uses a few additional components:

- *Rate modifier*: One per queue. This is either the bonus rate given to an active subscriber, or the substitute rate given to an inactive subscriber. Required in order for the assigned rate to be recovered.
- *Shared pool*: One only, within a control module that has access to all queues. This is the total combined rate currently being donated by inactive subscribers.
- *Contribution weight* or *just weight*: One per queue. This is the amount each subscriber's assigned rate contributes to the total assigned rate. For example, two subscribers are assigned 5 Mbps and 15 Mbps. The weight of the first subscriber is 0.25 (i.e., 5/20).

Inactive subscribers have their assigned rates removed from any bucket update calculations and their rate modifiers set to a small (configurable) fraction of their assigned rates, e.g., 10%. The remaining rate (90%) is donated to the shared pool. The shared pool is then distributed to active subscribers' rate modifiers based on their contribution weights. The exact formula used at this point can vary depending on the needs of the ISP. We will list and explain two possible examples:

**Balanced**: Active subscribers receive a rate of the shared pool multiplied by their contribution weight. This produces the most "fair" results - the maximum donated rate one can have

(i.e., all other subscribers are inactive) is exactly equal for all subscribers. If one subscriber upgrades their service, others have their contribution weight reduced, evening out the donation values. However, this method still produces a lot of discarded tokens; the total weight of all inactive subscribers is equal to the percentage wasted.

*Defined cap*: The entire shared pool is divided into all active subscribers based on their contribution weights, up to a maximum of their assigned rates multiplied by a configurable value. The typical amount of discarded tokens using this method should be very low; for example, if there are 10 subscribers each assigned 10 Mbps and a rate modifier cap of 2 (i.e., their maximum rate is double their usual rate), up to 5 of those subscribers can go inactive before the active subscribers reach their cap and tokens start being discarded. The rest of this paper will assume that this second method is used.

## C. Bucket Size

In addition to proportional token distribution, bucket sizes are also proportional to a subscribers' assigned rate based on a configurable value. For example, if a subscriber is assigned 5 Mbps and the bucket size multiplier for this access link is 8 bits per bps, their token bucket size would be 5 MB. This is to ensure that, regardless of assigned rates, all subscribers will reach their full threshold in approximately the same length of time. It also adds a further upgrade incentive, as a higher assigned rate means a larger token bucket, i.e., more time transmitting at peak rate even without rate donations.

## D. Reasoning

While rates are being shared, as long as the total combined token generation rate in an access switch remains constant, the long-term average transfer rate will be equal to or less than the sum of all assigned rates. If implemented correctly, and under the right circumstances (see section III B.), the proposed system can achieve a very high token sharing efficiency (i.e., greatly reducing the amount of tokens wasted) while a minority of subscribers are inactive.

It may be possible to implement such a system without token buckets, but TBF is an ideal foundation to build upon, as the token bucket itself (with its maximum size) allows us to easily detect when a user has been transmitting at a low rate for specific length of time and also allows for more concrete rules and behavior when that occurs. Distributing tokens that are not required immediately can be done in a simple and consistent manner with no discernible effect on transmission delay for that subscriber.

TABLE I.     "DEFINED CAP" EXAMPLE 1

| Subscriber | Active | TGR |
|---|---|---|
| A | Yes | 8 Mbps |
| B | Yes | 10 Mbps |
| C | Yes | 5 Mbps |

TABLE II.     "DEFINED CAP" EXAMPLE 2

| Subscriber | Active | TGR |
|---|---|---|
| A | Yes | 13.538 Mbps (+5.53) |
| B | No | 1 Mbps (-9) |
| C | Yes | 8.4615 Mbps (+3.46) |

## E. Issues

One of the issues with the proposed CTBF involves fairness with respect to time. If two subscribers (i.e., S1 and S2) are inactive initially and then S1 begins downloading a large file, it can transmit at a peak rate for longer than usual because S2 is still inactive. However, if S2 begins the same transmission shortly after, it receives no benefit because S1 is already active and none to donate. This can potentially loop into itself (S1 effectively consumes less tokens than S2 for the same download size, so its burst length is higher than S2's after they both finish) giving S2 less of a benefit over an extended period of time simply because S1 started first.

To mitigate this problem, one possibility is to add a "shared bucket" to the controller that is filled by tokens discarded while all subscribers are inactive, which are then distributed to active subscribers proportionally based on the time gap between the start of their bursts. A thorough investigation of this time-related unfairness with simulation experiments is under study.

## IV. NUMERICAL EXAMPLES

To demonstrate the benefits of the proposed CTBF, we carried out simulation experiments based on OMNeT++/ INET with the network setup and user behavior models described in [7]. Each test is run for 3 hours in simulation time with data gathered after a 20-minute warm-up period. The gathered data include the delay between a user issuing an HTTP request and receiving the document, the throughput of FTP transmissions, and the average decodable frame rate of a streamed "Silence of the Lambs" video detailed in [1].

The existing virtual test bed involves a varying number of subscribers each with 5 hosts (i.e., subscribers). To demonstrate the effectiveness of CTBF, it must be possible for a subscriber to become inactive or else they won't donate their rate. For this reason, in these tests, the number of users per subscriber is reduced from 5 to 1, and the token bucket size (except for the token bucket size test) is set to small enough for a single FTP transmission from one user to empty the token bucket before the download finishes. The goal with these changes is to create a simulation where subscribers are both reaching an inactive state and experiencing bonus average rates alternately, without having to change FTP file size or program special *long burst* behavior, both of which would differ this test bed too much from the original for it to be a worthwhile comparison.

First, we tested with a 100Mbps line and peak rate, a low rate per subscriber (2 Mbps), a small token bucket multiplier (8 bits per bps, i.e., 2MB) with a varying number of subscribers from 2 to 50. Then, we repeated the test with token sharing switched on. Adding more subscribers should increase stability of bonus rates but also reduce the bonus rates overall due to the shared pool being divided more times. We are interested in seeing whether these two balance each other out and produce the same average performance.

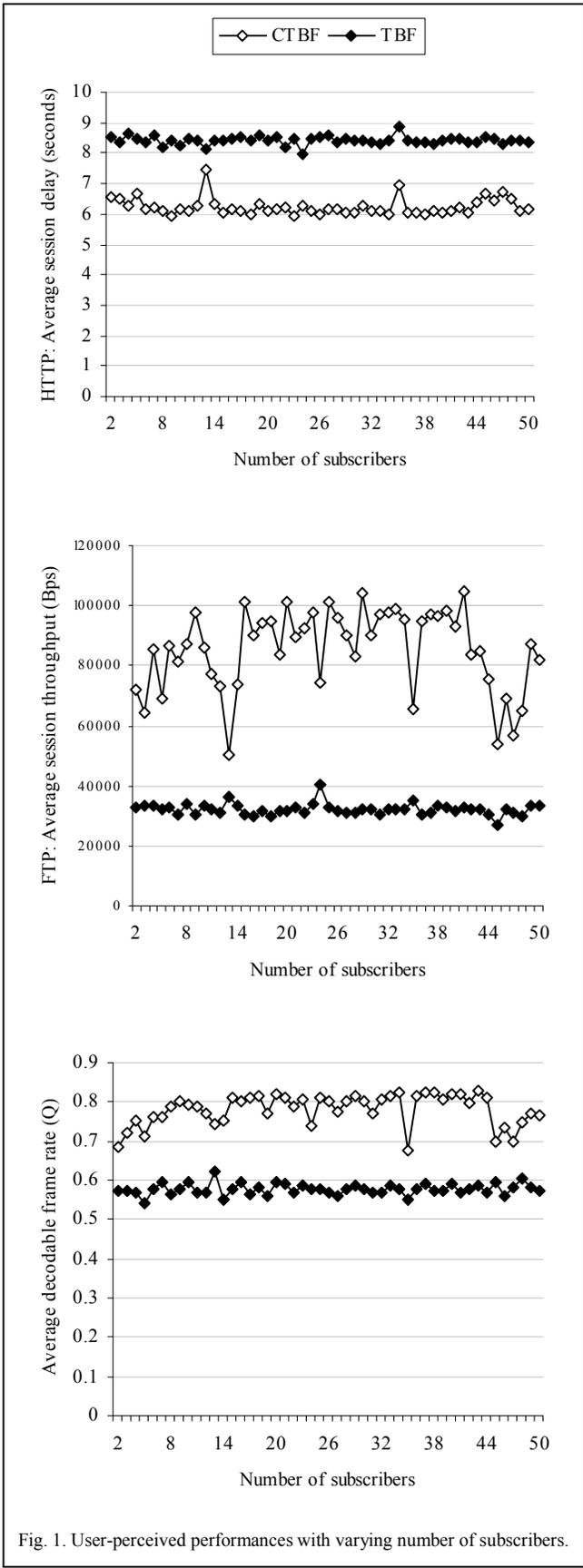

Fig. 1. User-perceived performances with varying number of subscribers.

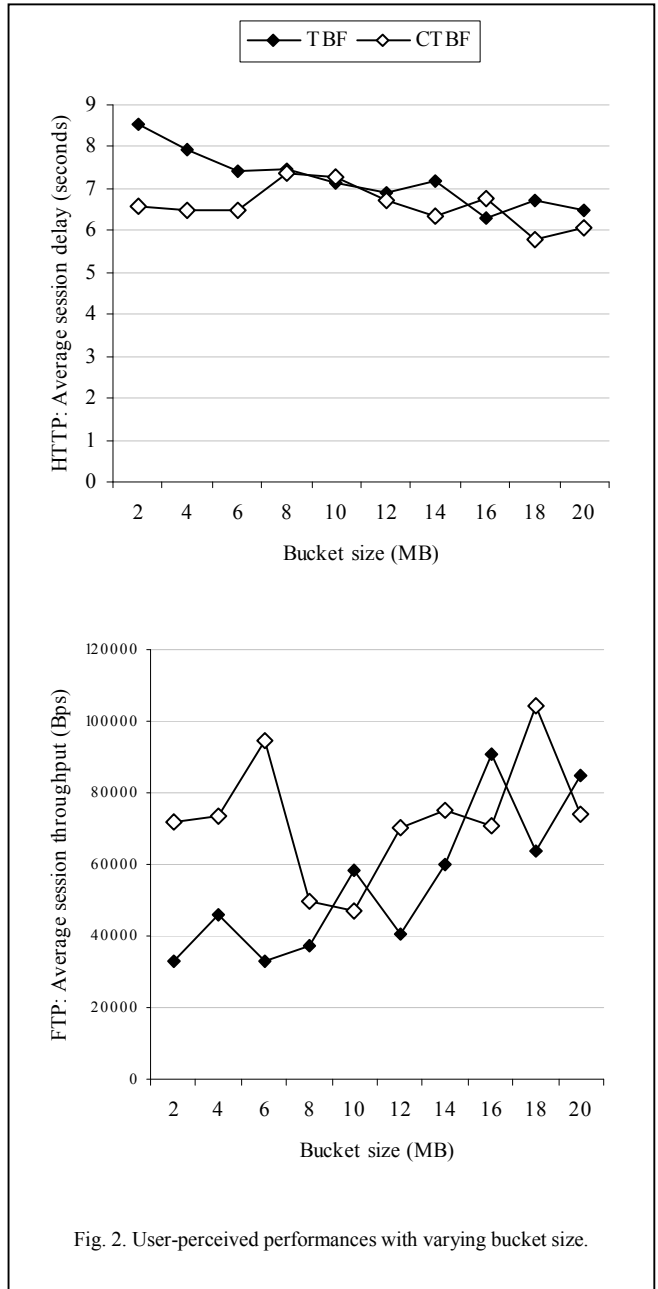

Fig. 2. User-perceived performances with varying bucket size.

Using the same rates and 2 subscribers only, we also tested with varying bucket sizes (i.e., 2MB to 20MB) to illustrate the potential conflict of CTBF and TBF. In TBF, the performance improvements with very large bucket sizes are astonishing, but in CTBF, as the bucket size increases, tokens may be shared with less frequency either due to subscribers not reaching their full thresholds or not downloading enough in a short space of time to experience the benefit of bonus average rates (as opposed to transmissions during bursts which are always the same rate). The latter is actually the best case scenario for both the schemes, as all subscribers are transmitting at no less than peak rate, given that the line rate allows it.

We expect the sharing results (especially FTP average session throughput) to be quite wild in comparison to the results without sharing; unlike with TBF whose results per subscriber are consistent, CTBF is much more sensitive to the behavior of individual users and therefore more susceptible to the randomness of the simulator. Regardless, the aforementioned stability with added subscribers should be visible in the results, and performance with sharing enabled should be noticeably superior in all categories.

In the subscriber test results shown in Fig. 1, the benefit of CTBF is very clear: Shorter average HTTP delays, and some improvement in the decodable frame rate of video streaming, but most notably a vast increase in the average FTP throughput: On some occasions, it is almost tripled due to users transmitting at the peak rate (i.e., 100Mbps) for longer period. The CTBF results vary wildly, but we can still make the following observation: Average bonus rates increase with the number of subscribers as the chance of some neighbors being inactive rises; then they gradually decline as the limit of the line rate is reached and congestion begins to occur. But even with the maximum number of subscribers (i.e., 50), the CTBF still shows advantages over the TBF.

In the bucket size tests shown in Fig. 2, we see that the CTBF produces consistently higher average rates and shorter average delays when the bucket size is small (i.e., file sizes are relatively large), but then begins to converge with the results from the TBF when the bucket size reaches 32 bits per bps (i.e., 8MB). The FTP file size is almost 5MB, and those transmissions can occur at the same time as video streaming (increasing usage by approximately 2Mbps), so this lines up with our prediction that the benefit of CTBF is dependent on whether subscribers download enough to empty their token buckets. But if they do not, they must be transmitting at a peak rate at all times which is the best case for user-perceived performance; CTBF is not necessary in such a case.

## V. CONCLUSIONS

In this paper we have investigated the TBF traffic shaping scheme and proposed an extended, theoretically non-inferior version that takes advantage of long-term burstiness of Internet traffic as well and is compatible with the ISP service contract model provided that the subscribers are made aware of how it works.

By recognizing that stored tokens are not required immediately, we found that the performance of active subscribers can be greatly improved without much affecting the QoS of inactive subscribers by sharing their discarded tokens only when their token bucket is full. Therefore we proposed the CTBF scheme which can take the benefits of excess bandwidth from those inactive subscribers. Through the preliminary simulation results presented in Section IV, we demonstrated that the CTBF produces the most consistent average performance with approximately 15 or more subscribers and is most effective when a subscriber unit is put under heavy load, e.g., file transmissions larger than the token bucket.

In this paper we haven't discussed the advantages of CTBF with more detailed simulation experiments, i.e., unique behavior of each subscriber, malicious subscribers, etc. These topics, along with the investigation of different types of packet scheduling after traffic shaping (e.g., round robin vs. longest wait first) are the subjects of a future study into CTBF.